\documentclass[preprint,showpacs,preprintnumbers,amsmath,amssymb]{revtex4}
\usepackage{graphicx}% Include figure files
\usepackage{dcolumn}% Align table columns on decimal point
\usepackage{bm}% bold math

\begin{document}

%\preprint{APS/123-QED}
\title{Leptonic and charged kaon decay modes of the $\phi$ meson measured in heavy-ion collisions at the CERN SPS}

\author{D.~Adamov{\'a} $^a$, G.~Agakichiev$^b$, D.~Anto{\'n}czyk$^b$,
 H.~Appelsh{\"a}user$^c$, V.~Belaga$^d$, J.~Biel\v{c}\'{\i}kov{\'a}$^e$$^k$,
 P.~Braun-Munzinger$^b$, O.~Busch$^b$, A.~Cherlin$^f$,
 S.~Damjanovic$^e$, T.~Dietel$^e$, L.~Dietrich$^e$, A.~Drees$^g$, S.~I.~Esumi$^e$, K.~Filimonov$^e$,
 K.~Fomenko$^d$, Z.~Fraenkel$^f$, C.~Garabatos$^b$, P.~Gl{\"a}ssel$^e$,
 G.~Hering$^b$, J.~Holeczek$^b$, V.~Kushpil$^a$, W.~Ludolphs$^e$, A.~Maas$^b$,
 A.~Mar\'{\i}n$^b$, J.~Milo\v{s}evi{\'c}$^e$, D.~Mi{\'s}kowiec$^b$, R.~Ortega$^e$,
Y.~Panebrattsev$^d$, O.~Petchenova$^d$, V.~Petr{\'a}\v{c}ek$^e$,
 S.~Radomski$^b$, J.~Rak$^b$, I.~Ravinovich$^f$, P.~Rehak$^h$, H.~Sako$^b$, W.~Schmitz$^e$,
J.~Schukraft$^i$, S.~Sedykh$^b$, S.~Shimansky$^d$, J.~Stachel$^e$,
M.~\v{S}umbera$^a$, H.~Tilsner$^e$, I.~Tserruya$^f$, G.~Tsiledakis$^b$,
J.\thinspace P.~Wessels$^j$, T.~Wienold$^e$, J.\thinspace P.~Wurm$^k$,
S.~Yurevich$^e$, V.~Yurevich$^d$\\
(CERES Collaboration)}

\affiliation{$^a$ NPI/ASCR, \v{R}e\v{z}, Czech Republic \\
$^b$ GSI, Darmstadt, Germany\\
$^c$ Universit{\"a}t Frankfurt, Germany\\
$^d$ JINR, Dubna, Russia\\
$^e$ Universit{\"a}t Heidelberg, Germany\\
$^f$ Weizmann Institute, Rehovot, Israel\\
$^g$ Department for Physics and Astronomy, SUNY Stony Brook, USA\\
$^h$ Brookhaven National Laboratory, Upton, USA\\
$^i$ CERN, Geneva, Switzerland\\
$^j$ Universit{\"a}t M{\"u}nster, Germany\\
$^k$ Max-Planck-Institut f{\"u}r Kernphysik, Heidelberg, Germany}

\date{\today}

\begin{abstract}
We report a measurement of $\phi$ meson production in central Pb+Au collisions at E$_{lab}$/A=158~GeV.
For the first time in heavy-ion collisions, $\phi$ mesons were reconstructed in the same experiment 
both in the K$^+$K$^-$ and the dilepton decay channel. Near mid-rapidity, this yields rapidity densities, 
corrected for production at the same rapidity value, 
of 2.05$\pm$0.14(stat)$\pm$0.25(syst) and 2.04$\pm$0.49(stat)$\pm${0.32}(syst), respectively. The shape of the
measured transverse momentum spectra is also in close agreement in both decay channels. 
The data rule out a possible enhancement 
of the $\phi$ yield in the leptonic over the hadronic channel by a factor larger than 1.6 at 95$\%$ CL.

\end{abstract}

                             
                              
\maketitle

%\section{Introduction}

In ultrarelativistic nucleus-nucleus collisions matter is created at high temperature and density. For top 
SPS energies and higher, the corresponding fireball is very likely \cite{qm04all} in a deconfined state as 
predicted by solutions of Quantum Chromodynamics (QCD) on a space-time lattice \cite{lattice}. 
Detailed studies of hadron 
abundances imply \cite{pbm} that their yield is frozen very close to the predicted phase boundary between 
confined and deconfined matter. 
Strangeness enhancement has been suggested \cite{rafelski} as a signature of a deconfined stage. 
In this context it is important to understand the production of 
$\phi$ mesons which carry hidden  strangeness. Furthermore, near the phase boundary the $\phi$ meson mass, 
width and its branching ratios into kaons and leptons might be modified \cite{weise,shuryak}. Due to final state 
interactions of the kaons, such a modification would probably only be visible in the lepton decay channel, 
which could result in different $\phi$ yields depending on which decay channel is studied \cite{joh}.
Indeed, the experimental situation concerning $\phi$ production at the top Super Proton Synchrotron (SPS) 
energy provides hints for such a scenario as the NA50 collaboration reported 
a $\phi$ yield measured via dileptons \cite{na50} which exceeded the yield determined by NA49 
in the K$^+$K$^-$ channel \cite{Friese} by factors between two and four \cite{puzzle}
in the common transverse momentum range.
 Further, the $m_t$ spectra exhibit a different inverse
slope parameter \cite{puzzle}, 305~$\pm$~15 MeV in NA49 and 218~$\pm$~6 MeV in NA50,
fitted in their $m_t$ acceptance regions.

The upgrade of the CERES experiment \cite{up2,up1,qm99,qm04} makes possible for the first time to 
simultaneously study the leptonic and the charged kaon decay modes of
the $\phi$ meson at the SPS, thus shedding light onto the $\phi$ puzzle.
In this paper, we present results of $\phi$ mesons reconstructed both 
in the charged kaon (K$^+$K$^-$) and in the dilepton (e$^+$e$^-$) decay mode.
The  kaon (dilepton) analysis uses 24 (18) million 158 AGeV/c Pb on Au collisions taken at the most 
central 7\% of the geometrical cross section.
%The data taking rate was 300 to 500 events per spill (of 4.2 s duration and 
%of 19.2 s cycle time) at a Pb beam intensity of 10$^6$ particles per spill.

The CERES experiment is optimized to measure low mass electron pairs close to
mid-rapidity (2.1$<\eta<$2.65) with full azimuthal coverage
\cite{pp,exp,nantes,9596,40gev}. A vertex telescope, composed of two Silicon
Drift Detectors (SDD) positioned at 10 cm and 13.8 cm downstream of a
segmented Au target (thickness equivalent to 1.33\% interaction
length), provides a precise vertex reconstruction.
Two Ring Imaging CHerenkov (RICH) detectors,
operated at a high threshold ($\gamma_{th}$=32), provide electron
identification in a large hadronic background. 
The new radial-drift TPC, positioned downstream of the original 
spectrometer inside a magnetic field provides the momentum of 
charged particles with a resolution $\Delta p/p\sim((2\%)^2+(1\%\cdot p$(GeV/c))$^2)^{1/2}$ 
and additional electron identification via d$E$/d$x$.
Charged particles from the target are reconstructed by matching track segments in the 
SDD and in the TPC using a momentum-dependent matching window. 
Matching to a ring in the RICH is also required for the dilepton analysis. 

To study the $\phi$ meson in the charged kaon (K$^+$K$^-$) decay mode, all charged
particles get assigned the kaon mass (no particle identification is used). 
Only a conservative upper cut in the d$E$/d$x$ signal (corresponding to 90\% of the 
Fermi plateau value) for momenta between 1.25 GeV/c and 4 GeV/c, suppressing 83\% of the 
electrons, is applied to enhance the kaon content of the sample.
Tracks in the TPC are required to contain more than 12 hits of maximum 20 possible to ensure good
momentum resolution. Tracks in the geometrical acceptance 0.13~rad
$<\theta<$~0.24~rad with a transverse momentum $p_t$ larger than 0.25~GeV/c are selected.
To reduce the contamination from other particle species, cuts in the Podolanski-Armenteros 
parameter \cite{armen} and in the opening angle between the kaons are applied \cite{longpaper}.
The $\phi$ meson in the kaon decay mode is studied in the rapidity
interval 2.0~$<y^\phi<$~2.4 for $p_t^\phi>0.75$~GeV/c.
%given by the acceptance of the experiment.

\begin{figure}
\includegraphics[width=9.5cm]{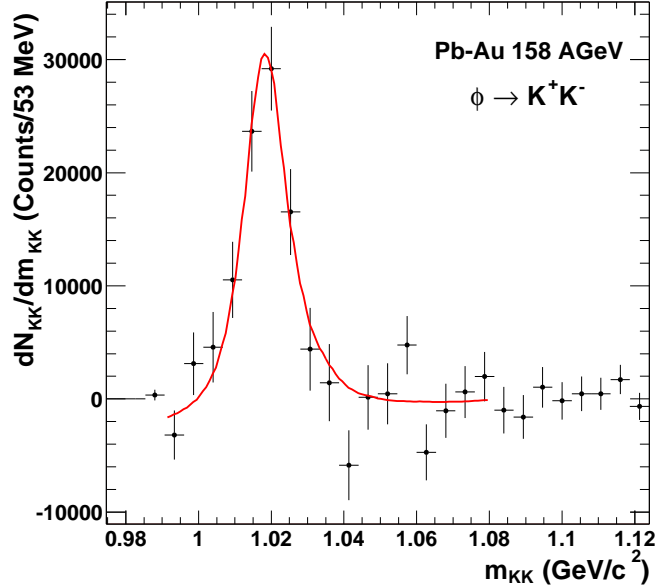}
\caption{Invariant-mass spectra of $K^+K^-$ pairs after background
subtraction for the ($p_t^\phi$,$y^\phi$) bin 1.5 GeV/c $<$ $p_t^\phi$ $<$ 1.75 GeV/c
and 2.2 $<$ $y^\phi$ $<$ 2.4.}
\label{fig:phi_ka}
\end{figure}

To obtain the $\phi$ transverse momentum spectrum the invariant mass
distributions of K$^+$K$^-$ pairs were
accumulated in ($p_t^\phi$,$y^\phi$) bins.
The invariant mass distributions of the combinatorial background are calculated
using the mixed-event technique for each ($p_t^\phi$,$y^\phi$) bin.
One such invariant mass distribution, after background subtraction, 
is presented in Fig.~\ref{fig:phi_ka}.
Residual background appears in the low $p_t^\phi$ bins. The yield of 
the $\phi$ mesons is, therefore, determined by fitting a relativistic 
Breit-Wigner distribution with parameters taken from the Particle Data Group compilation \cite{pdg}
(convoluted with the experimental resolution function obtained by a Monte Carlo simulation)
superimposed on a linear background, to the measured line shape.
 The signal to background ratios vary from 1/2000 to 1/180 with increasing $p^\phi_t$.
The signal is integrated in the mass range between 1.0 GeV/c$^2$ and 1.05 GeV/c$^2$.
The resulting $\phi$ yields need to be corrected for acceptance and efficiency. 

To study the $\phi$ meson in the dilepton (e$^+$e$^-$) decay mode, electrons 
are identified among all charged hadrons
using the RICH detectors and the d$E$/d$x$ signal in the TPC.
The two RICH detectors are used in a combined mode. 
Cherenkov rings with asymptotic radius are identified using a Hough 
transformation (see sect. 3.2.4 in ref. \cite{9596}).
A pion rejection factor of 2000 is achieved in the RICH \cite{busch} for an
electron efficiency of 0.70 with the quality cuts applied in the analysis \cite{serguei,alex}.
The TPC electron selection is done based on the d$E$/d$x$ signal 
and its resolution.
The combined pion rejection factor varies from 4$\times$10$^4$ to 1.8$\times$10$^4$ for momenta 
between 1~GeV/c and 2.5~GeV/c for total electron efficiencies of 68\% and 66\% \cite{serguei}, respectively.

%Electrons are reconstructed by combining track segments
%in SDD, TPC and RICH that fulfill certain quality criteria with a 
%2$\sigma$ momentum-dependent matching window \cite{serguei,alex}. 

The main difficulties of reconstructing $\phi$ mesons in the di-electron channel 
are the low branching ratio of this decay mode and the large amount of 
combinatorial background from $\gamma$ conversions and Dalitz decays of neutral mesons. Therefore,
besides a very good electron identification, removal of 
electron pairs from $\gamma$ conversions and $\pi^0$ Dalitz decays from the sample 
is vital to reduce the combinatorial background. 
The details of the rejection strategy are explained in \cite{qm04,qm05,serguei,alex}.
Electron tracks in the geometrical
acceptance 0.14~rad~$< \theta< $~0.243 rad and with
a transverse momentum $p_t>$~0.2~GeV/c are selected \cite{serguei,alex}.
The $\phi$ meson in the dilepton decay mode is studied in the rapidity interval 2.1~$<$~$y^\phi$~$<$~2.65.
The signal in the e$^+$e$^-$ channel is obtained from 
the invariant-mass distributions of unlike-sign pairs after full rejection and 
subtraction of the mixed-event combinatorial background. The mixed-event background is
normalized to the like-sign pair background in the mass region 0.2~GeV/c$^2<m_{e^+e^-}<$1.6~GeV/c$^2$.
The integrated yield in the mass range between 0.9 and 1.1 GeV/c$^2$ is 229$\pm$53 with 
a signal to background ratio of 1/12 and 
needs to be corrected for acceptance, reconstruction efficiency and physics background under the $\phi$ peak.

In order to determine the acceptance, decay and efficiency corrections, 
Geant \cite{geant} simulations containing the description of the CERES experiment are used.
The Monte Carlo simulation is tuned to reproduce all aspects of the data \cite{qm04,serguei,alex}. 
Sources of $\phi$ mesons with realistic transverse momentum and rapidity distributions 
were embedded into real events to simulate the background. 
The simulation of the dilepton channel is done using the GENESIS simulation code \cite{cocktail}.
All ($p_t^\phi$,$y^\phi$) bins with an acceptance larger than 0.5 are used in the K$^+$K$^-$ analysis.  
The reconstruction efficiency is of the order of 0.4 with slight dependence on the transverse
momentum, given by the decay in flight of the charged kaons, the pair cuts applied and the 
single track efficiency.
The acceptance for the e$^+$e$^-$ pairs varies from 0.165 to 0.185 in the measured rapidity range 
and the reconstruction efficiency after full rejection varies from 0.22 to 0.145 with the $\phi$ 
transverse momentum.
An additional loss of 25\% is due to bremsstrahlung events that lead to a reduction of 
the mass by more than 100 MeV.

\begin{figure}[htb]
\includegraphics[width=9.cm]{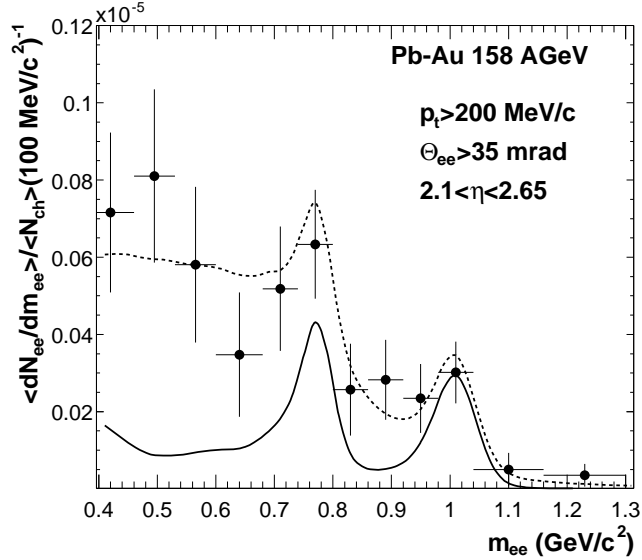}
%\vspace{-1cm}
\caption{Invariant mass spectrum of e$^+$e$^-$ pairs per charged particle compared to 
the hadron decay cocktail (thick solid line). Also plotted is a model calculation assuming 
in medium spread $\rho$ width plus the dilepton yield from the QGP phase (dashed line) used to extract the
physics background in the $\phi$ mass region.}
\label{fig:maselec}
\end{figure}

The e$^+$e$^-$ invariant-mass spectrum corrected for 
efficiency and normalized to the number of charged particles 
in the acceptance \cite{qm05} is shown in Fig.~\ref{fig:maselec} 
%for masses $m_{e^+e^-}>$0.4~GeV/c$^2$ 
together with the expectations from the hadron decay cocktail
\cite{cocktail}. 
%Acceptance, opening-angle, and transverse-momentum cuts are applied.
The $\phi$ meson yield in the e$^+$e$^-$ channel is determined by integrating the
invariant mass spectra in the mass region between 0.9 and 1.1 GeV/c$^2$ 
in three transverse momentum bins. The $\rho$ meson could extend into this mass range if 
its spectral function is modified in the medium. 
Dileptons from the QGP phase also contribute to the physics background in this mass range.
The sum of these two contributions is estimated to be 35\% 
of the total yield in this mass region by inspecting theoretical models
that include in-medium spreading of the $\rho$ width due to 2$\pi$ processes and the dilepton yield from the QGP phase \cite{theo1}.
If processes  involving 4 or 6 pions would be included in \cite{theo1} the physics background could be larger.
Using another model \cite{theo2} that also describes the CERES data gives a physics background  
contribution of 37\%. The measured $\phi$ yield has been scaled by the  
smaller factor to correct for the physics background.
The charm contribution that is smaller than 3\% \cite{charm} has been neglected.

\begin{figure}[htb]
\includegraphics[width=9.5cm]{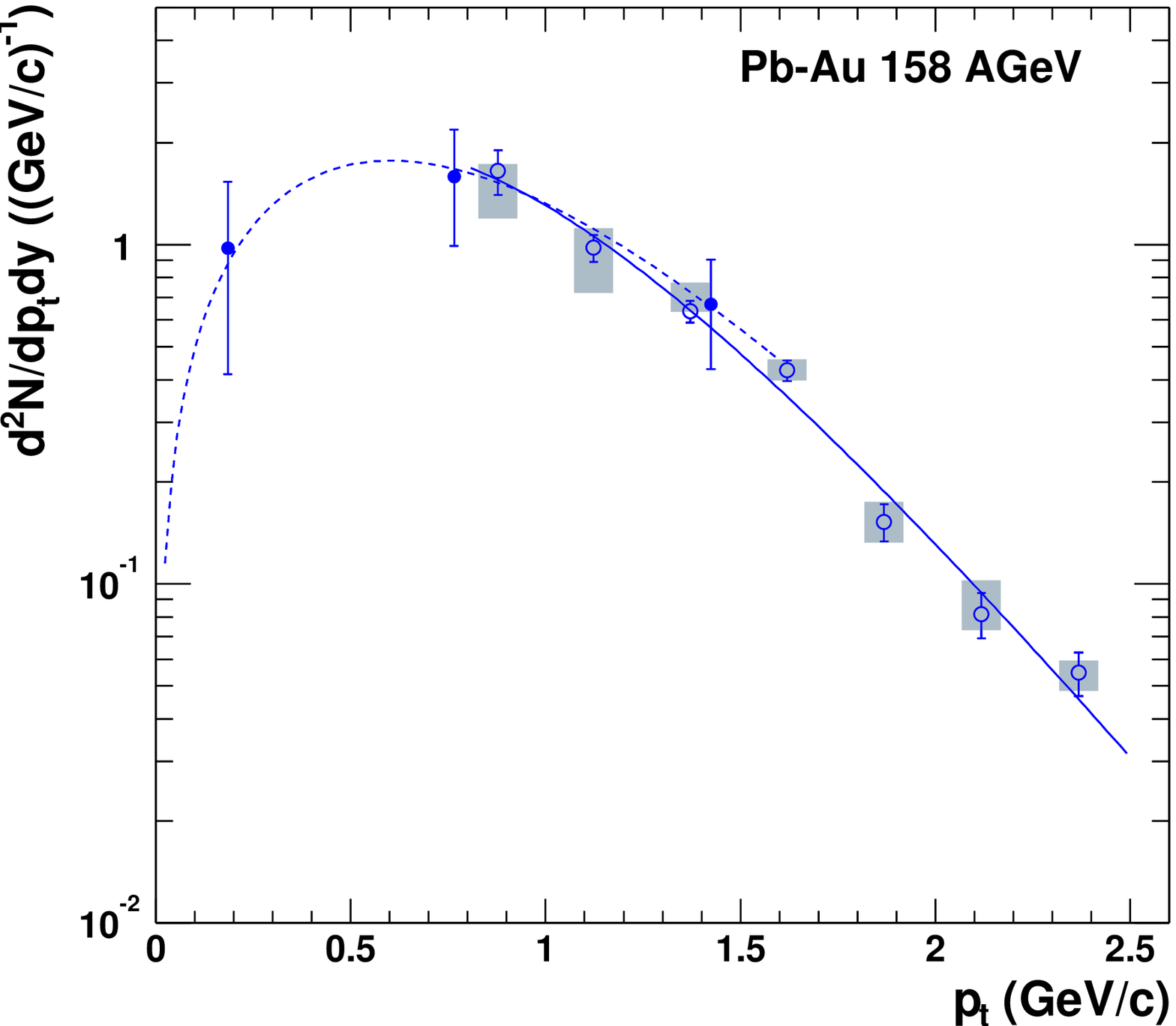}
\caption{Transverse momentum spectrum of $\phi$ mesons corrected for
acceptance and efficiency reconstructed in the e$^+$e$^-$ decay 
mode (closed symbols) and in the K$^+$ K$^-$ decay channels (open symbols). 
The spectrum in the e$^+$e$^-$ decay mode has been scaled by 0.93 before comparing to the
K$^+$ K$^-$ decay channels to account for the different rapidity intervals covered in the two analyses.
Systematic errors in the kaon decay channel are shown as boxes.}
\label{fig:ptspec}
\end{figure}

The efficiency- and acceptance-corrected $\phi$ meson yield is shown in Fig.~\ref{fig:ptspec} for both decay 
modes as a function of transverse momentum. 
The error bars shown are statistical.
The systematic errors in the charged kaon analysis (plotted as boxes) contain contributions 
from variation of the pair cuts, from the difference between the integral from the fit and 
from data points in the given interval, from the dependence of the yield on the fit window, and 
the function used to fit the residual background. The systematic errors in the dilepton analysis
are $\pm$16\%. They contain the contribution from the analysis cuts, background substraction and 
variation of mass range. They are not drawn because 
the errors in the dilepton channel are totally dominated by statistics.

When the spectra are fitted with the function
$$d^2N/dp_tdy={{dN/dy}\over{T\cdot(T+m_\phi)}}\cdot p_t \cdot \exp{(-{{m_t-m_\phi}\over{T}})}$$
an inverse slope parameter of $T$~=~273~$\pm$~9(stat)$\pm$10(sys) MeV and a rapidity density 
$dN/dy$ of 2.05~$\pm$~0.14(stat)$\pm$~0.25(sys) in the K$^+$K$^-$ decay mode 
and $T$~=~306~$\pm$~82 MeV and $dN/dy$~=~2.19~$\pm$~0.52(stat)$\pm${0.34}(syst) in the dilepton decay 
mode are obtained averaging over their corresponding rapidity intervals. 
In order to compare the rapidity densities in the two decay channels
the dilepton data are scaled by \cite{Friese} (d$N_{\phi}$/d$y$)$^{2-2.4}$ =0.93$\cdot$(d$N_{\phi}$/d$y$)$^{2.1-2.65}$.    
The $\phi$ meson yields and inverse slope parameters obtained in both decay modes  
agree within the errors. A $\phi$ yield in the e$^+$e$^-$ decay mode larger 
than 1.6 times the yield on the K$^+$K$^-$ decay mode is excluded at 95\% CL 
(statistical and systematic errors in both decay channels added in quadrature).

Moreover, the CERES results can be compared to the existing Pb-Pb systematics \cite{puzzle} 
after accounting for the different measurement conditions.
 The NA49 measurement was done at 4\% centrality and covered a rapidity range from 3 to 3.8
units \cite{Friese}. 
\begin{figure}[hbt]
%\vspace*{-0.2cm}
\includegraphics[width=9.5cm]{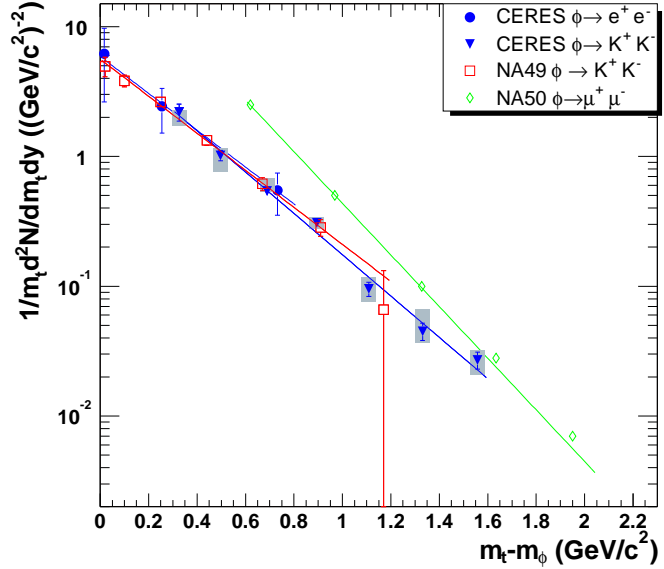}
\caption{Transverse mass distribution of $\phi$ mesons measured in the
charged kaon (triangles) and in the dilepton (circles) decay mode 
after scaling (see text) compared to the results from NA49 (squares) and NA50 (diamons). 
Error bars are statistical and systematic errors in the kaon channel are shown as boxes.}
\label{fig:mtcomp}
\vspace*{0.5cm}
\end{figure}
%The increase of $<\phi>$/$<\pi>$  with N$_{part}$ is neglected,
%since the two centralities are very similar.
%On the other hand 
We use the measured increase of charged hadron (h$^-$) multiplicity with 
centrality ( ${{h^{-}_{4\%}}/{h^{-}_{7\%}}}$ = 1.092$\pm$0.014 ) \cite{longpaper} to correct
 the $\phi$ meson yield  for the different centralities between the two experiments.
From the NA49 $\phi$ meson rapidity distribution a ratio of
(dN$_{\phi}$/dy)$^{3-3.8}$/(dN$_{\phi}$/dy)$^{2.-2.4}$  = 1.07 $\pm$ 0.11   
is obtained to account for the different rapidity coverage.
Thus, a global scaling factor of $1.17 \pm 0.12$
is applied to the combined CERES data of Fig.~\ref{fig:ptspec} to make the comparison 
to the systematics of \cite{puzzle}.
In Fig.~\ref{fig:mtcomp} the $\phi$ transverse mass spectrum obtained
 by CERES after scaling is plotted together with the NA49 and NA50 data.
The $\phi$ meson yields agree within the errors with the NA49 results. So does 
the yield in the K$^+$K$^-$ extrapolated down to $p_t$=0 using the measured inverse 
slope parameter.
On the other hand, CERES data in the K$^+$K$^-$ channel do not agree with NA50 results in
the common $p_t$ region. This experiment however measures the leptonic channel. 
The extrapolation of NA50 results down to the region where CERES measures the 
dilepton channel does not agree either. As stated above, in the CERES measurement the 
two decay modes agree. Possible differences of maximum 40-50\% as expected by models including 
only rescattering of the kaons \cite{joh} or of maximum 70\% at the 
lowest $p_t$ ($p_t < 0.3$ GeV/c) expected by 
models including medium modifications of the $\phi$ mesons and kaons like the 
AMPT model \cite{pal02} cannot be ruled out by the CERES results. 

To conclude, $\phi$ meson production has been measured simultaneously in both decay channels 
for the first time at the CERN SPS. The yield and the inverse slope parameter in both 
decay modes agree within the errors. 
Our results are in agreement with the results from NA49 measured in the kaon channel. 
A yield in the e$^+$e$^-$ decay mode larger 
than 1.6 times the yield on the K$^+$K$^-$ is excluded at 95\% CL, therefore the large discrepancy 
observed previously is not observed in the CERES data.
The theoretical predictions in \cite{joh,pal02} are consistent with our data but with  
our precision we cannot distinguish between the different effects.

The CERES collaboration acknowledges the good performance of the CERN
PS and SPS accelerators as well as the support from the EST
division. We would like to thank R.~Campagnolo, L.~Musa, A.~Przybyla, W.~Seipp and B.~Windelband for their 
contribution during construction and commissioning of the TPC and during data taking.
 We are grateful for excellent support by the CERN IT division for the central data
recording and data proccesing. This work was supported by GSI, Darmstadt, the German BMBF, the German VH-VI 146, the US DoE, the Israeli Science
Foundation, and the MINERVA Foundation.


\begin{thebibliography}{99}

\bibitem{qm04all} See, e.g., Proc. Quark Matter 2004, {\it J. Phys. G: Nucl. Part. Phys. 30}.

\bibitem{lattice} F. Karsch, {\it J. Phys. G: Nucl. Part. Phys.} 31 (2005) S633. 
F. Karsch, {\it Nucl. Phys.} A 698 (2002) 199c.

\bibitem{pbm} P. Braun-Munzinger, J. Stachel, C. Wetterich, {\it Phys. Lett.} B 596 (2004) 61.

\bibitem{rafelski} P.~Koch, B.~M{\"u}ller, J.~Rafelski {\it Phys. Rep.} 142 (1986) 167.

\bibitem{shuryak} D. Lissauer and E. V. Shuryak, {\it Phys. Lett.} B 253 (1991) 15.

\bibitem{weise} F. Klingl, T. Waas, W. Weise, {\it Phys. Lett.} B 431 (1998) 254. 

\bibitem{joh} S.C.~Johnson, B.V.~Jacak, A.~Drees, {\it Eur. Phys. J. C}18 (2001) 645.

%\bibitem{koc1} P.~Koch, B.~M{\"u}ller, J.~Rafelski {\it Phys. Rep. 142}(1986) 167.
%\bibitem{koc2} V.~Koch {\it Int. J. Mod. Phys. E} (1997) 203; F.~Karsch{Nucl. Phys. A 590} (1995) 367c.


\bibitem{na50} N.~Willis {\it et al.}, NA50 Collaboration, {\it Nucl. Phys.} A 661 (1999) 543c; 
B. Alessandro {\it et al.}, {\it Phys. Lett.} B 555 (2003) 147.

\bibitem{puzzle} D.~R{\"o}hrich, {\it J. Phys. G } 27 (2001) 355.

\bibitem{Friese} S.V.~Afanasiev {\it et al.}, NA49 Collaboration {\it Phys. Lett. B} 491 (2000) 59.

\bibitem{up2} Addendum to proposal SPSLC/P280: CERN/SPSLC 96-35/P280 Add.1.

\bibitem{up1} Technical Note on the NA45/CERES upgrade. CERN/SPSLC 96-50 (1996).

\bibitem{qm99} A. Mar\'{\i}n for the CERES Collaboration, {\it Nucl. Phys.} A 661 (1999) 673c.

\bibitem{qm04} A. Mar\'{\i}n {\it et al.}, CERES Collaboration, {\it
J. Phys. G: Nucl. Part. Phys. 30} (2004) S709.

\bibitem{pp} G.~Agakichiev {\it et al.}, CERES Collaboration, {\it
Eur. Phys. J.} C 4 (1998) 231.

\bibitem{exp} G.~Agakichiev {\it et al.}, CERES Collaboration, {\it Phys. Rev. Lett.}
75 (1995) 1272.

\bibitem{nantes} G.~Agakichiev {\it et al.}, CERES Collaboration, {\it
Phys. Lett. B} 422 (1998) 405; B.~Lenkeit for the  CERES Collaboration,
{\it Nucl. Phys.} A 661 (1999) 23c; J.P.~Wessels for the  CERES
Collaboration, {\it Nucl. Phys.} A 715 (2003) 607c.

\bibitem{9596} G.~Agakichiev {\it et al.}, CERES Collaboration, {\it Eur. Phys. J.} C 41 (2005) 475.

\bibitem{40gev} D.~Adamova {\it et al.}, CERES Collaboration, {\it  Phys. Rev. Lett.} 91 (2003) 042301.

\bibitem{armen} J.~Podolanski and R. Armenteros, {\it Phil. Mag.} 45 (1954) 13.

\bibitem{longpaper} CERES Collaboration, {\it Reconstruction of $\phi$ mesons in the K$^+$K$^-$ channel}, 
in preparation.

\bibitem{pdg} S.~Eidelmal {\it et al.}, {\it Phys. Lett.} B 592 (2004) 1.

\bibitem{busch} O.~Busch, Doctoral Thesis, in preparation, TU Darmstadt (2005).

\bibitem{serguei} S.~Yurevich, Doctoral Thesis, in preparation, University of Heidelberg (2005).

\bibitem{alex} A.~Cherlin, Doctoral Thesis, in preparation, Weizmann Institute of Science (2005).

\bibitem{geant} GEANT, Detector Description and Simulation Tool, CERN Program Library Long Writeup W5013.

\bibitem{qm05} D.~Mi{\'s}kowiec for the CERES Collaboration. Proc. Quark Matter 2005 {\it Nucl. Phys.} A in print, nucl-ex/0511010.

\bibitem{cocktail} H.~Sako for the CERES Collaboration. Technical Report 03-25 (2000).

\bibitem{Friese02} V.~Friese for the NA49 Collaboration, {\it Nucl. Phys. A} 698 (2002) 487c.

\bibitem{sik} F. Sikler for the NA49 Collaboration, {\it Nucl. Phys. A} 661 (1999) 45c.


\bibitem{theo1} R.~Rapp and J. Wambach, {\it Adv. Nucl. Phys.} 25 (2000) 1; R.~Rapp, private comunication; 
E.~Braaten, R.D.~Pisarski and T.-C. Yuan, {\it Phys. Rev. Lett.} 64 (1990) 2242.

\bibitem{theo2} K. Gallmeister {\it et al.}, {\it Nucl. Phys.} A 688 (2001) 939; and B. K{\"a}mpfer, private communication.

\bibitem{charm} P.~Braun-Munzinger, D.~Mi{\'s}kowiec, A.~Drees, C.~Louren\c{c}o, {\it Eur. Phys. J. C} 1 (1998) 123.

\bibitem{pal02} S.~Pal, C.M.~Ko, Zi-wei~Lin, {\it Nucl. Phys.} A 707 (2002) 529.

\end{thebibliography}
\end{document}